\documentclass[osajnl,twocolumn,showpacs,superscriptaddress,10pt]{revtex4-1} 
\usepackage{amsmath,amssymb,graphicx}
\usepackage[utf8]{inputenc}
\usepackage[T1]{fontenc}
\usepackage{siunitx}
\usepackage{color}

\begin{document}

\title{On phase-locking of Kerr combs}

\author{Aur\'elien Coillet}\email{Corresponding author: aurelien.coillet@femto-st.fr}
\affiliation{FEMTO-ST Institute [CNRS UMR6174], Optics Department,
16 Route de Gray, 25030 Besan\c con cedex, France.}

\author{Yanne Chembo}
\affiliation{FEMTO-ST Institute [CNRS UMR6174], Optics Department,
16 Route de Gray, 25030 Besan\c con cedex, France.}

\begin{abstract}
We theoretically investigate the phase-locking phenomena between the spectral
components of Kerr optical frequency combs in the dynamical regime of Turing
patterns. We show that these Turing patterns display a particularly strong and
robust phase-locking, originating from a cascade of phase-locked triplets which
asymptotically lead to a global phase-locking between the modes. The local and
global phase-locking relationship defining the shape of the optical pulses are
analytically determined. Our analysis also shows that solitons display a much
weaker phase-locking which can be destroyed more easily than in the Turing
pattern regime. Our results indicate that Turing patterns are generally the most
suitable for applications requiring the highest stability. Experimental
generation of such combs is also discussed in detail, in excellent agreement
with the numerical simulations.    
\end{abstract}

\ocis{140.3945,140.4050,190.4380,190.5530.}

\maketitle 

Kerr optical frequency combs have been the focus of a strong interest from the
scientific community since their first generation a few years
ago~\cite{Kippenberg2011}. These combs result from a four-wave mixing process,
where new frequencies are generated through the nonlinear interaction between
the light and the bulk material of a monolithic whispering-gallery mode (WGM)
resonator. Depending on the frequency and power of the pump laser, this
phenomenology leads to the excitation of equally spaced spectral lines, thereby
forming the so-called Kerr comb. This multi-wavelength optical source is very
promising for various applications such as high-resolution spectroscopy and
ultra-stable microwave generation. In the latter case, the generated microwave
corresponds to the beating between the different spectral modes, and its phase
noise is directly related to the phase relationship between the comb's
lines~\cite{PRL_Nist,Vahala2012,Matsko2013}. The understanding of the
phase-locking process and its robustness with regard to external perturbations
is therefore of the highest importance for ultra-stable microwave generation,
and the objective of this letter is to unveil the mechanisms ensuring a stable
and robust phase-locking between the modes.

\begin{figure}[t]
  \centerline{
    \includegraphics{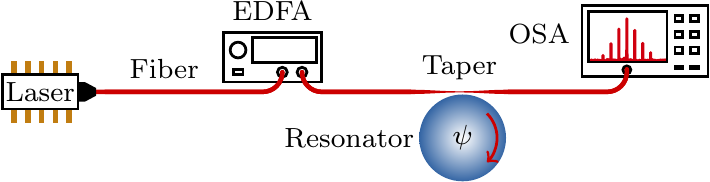}
   }
  \caption{Experimental setup.}
  \label{fig:Schema}
\end{figure}

The experimental setup used for Kerr comb generation is shown on
Fig.~\ref{fig:Schema}. The output of a continuous-wave (CW) $1550$~nm laser is
amplified and coupled to an ultra-high $Q$ WGM resonator using the evanescent
field of a tapered fiber. Above a given pump power threshold, a Kerr comb is
generated and can be monitored using an optical spectrum analyzer.

A spatiotemporal formalism based on the Lugiato-Lefever equation (LLE) has
recently been proposed to describe Kerr comb generation in WGM
resonators~\cite{Matsko2011,PRA_Yanne-Curtis,Coen2013}. In our case, the LLE
can be writtten as~\cite{PRA_Yanne-Curtis}:
\begin{equation}
  \frac{\partial \psi}{\partial \tau} = - (1 + i \alpha)\psi + i |\psi|^2 \psi -
  i \frac{\beta}{2}\frac{\partial^2\psi}{\partial \theta^2} + F 
  \label{eq:LLE}
\end{equation}
where $\psi (\theta, \tau)$ is the complex amplitude of the overall field in the
cavity, $\theta \in [- \pi, \pi]$ is the azimuthal angle along the cavity
circumference, $\tau = t/2 \tau_{\rm ph}$ is the normalized time, with
$\tau_{\rm ph} = 1/ \Delta \omega_{\rm tot}$ being the photon lifetime, $\Delta
\omega_{\rm tot} = \omega_0/Q_{\rm tot}$ the loaded linewidth, and $Q_{\rm tot}$
the loaded quality factor. The parameter $\alpha = - 2(\Omega - \omega_0) /
\Delta \omega$ stands for the detuning between the pump laser  and the cold
resonance frequencies $\Omega$ and $\omega_0$, respectively. The parameter
$\beta$ represents the second-order dispersion of the resonator, while $F$
stands for the amplitude of the pump laser field.

\begin{figure}
  \centerline{
    \includegraphics{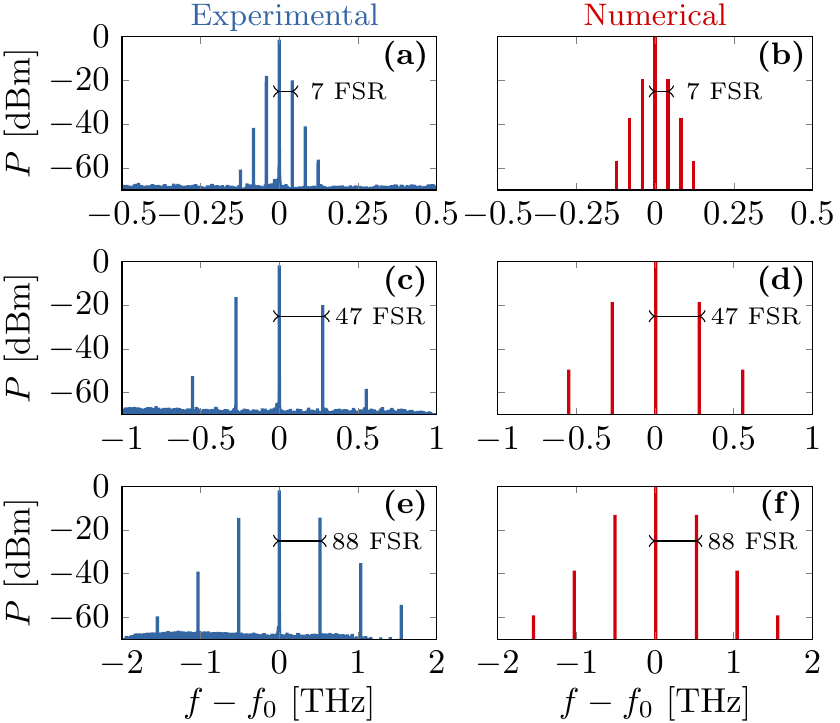}
  }
  \caption{Comparison between experimental (a, c and e) and theoretical (b, d
  and f) multiple-FSR spectra of Kerr combs corresponding to Turing patterns. 
  Various line spacings are observed ranging from 7 to 88~FSR ($41$ to $516$~GHz).}
  \label{fig:CompExpSim}
\end{figure}

Various non-trivial solutions can emerge from the LLE depending on the
parameters $\alpha$, $\beta$ and $F$~\cite{Balakireva2013,Godey2013}.  In the
regime of anomalous dispersion, the time-independent stationary solutions are
either Turing patterns or bright solitons.  In the spatio-temporal description,
Turing patterns correspond to an integer number of intensity extrema along the
azimuthal direction of the cavity.  In the Fourier domain, these Turing patterns
correspond to the well-known Kerr combs with multiple free-spectral range (FSR)
spacing~\cite{Savchenkov2008,YanneNanPRL,YanneNanPRA,Coillet2013}. 

A few examples of such combs are presented in Fig.~\ref{fig:CompExpSim}.  They
have been generated in a magnesium fluoride (MgF$_2$) resonator with intrinsic
$Q$-factor equal to $1.5 \times 10^9$, and a FSR equal to $\Omega_{\rm FSR}/2
\pi = 5.86$~GHz.  By choosing different resonances, pump powers and frequencies,
we were able to obtain different mode spacings, each corresponding to a
different integer number of FSR~\cite{Savchenkov2008}.  For each one of these
spectra, numerical simulations were performed using the split-step Fourier
method on the LLE, and a good agreement with the experiments was found, hence
validating the theoretical description.  In the case of
Fig.~\ref{fig:CompExpSim}(b), the parameters used were $\beta=-0.03$,
$\alpha=1.27$ and $F^2=1.075$. For Fig.~\ref{fig:CompExpSim}(d), $\beta$ is
equal to $-0.003$, $\alpha=-1.3$ and $F^2=6.4$, while in
Fig.~\ref{fig:CompExpSim}(f), $\beta=-0.001$, $\alpha=-1.87$ and $F^2=14$. 

\begin{figure}
    \includegraphics{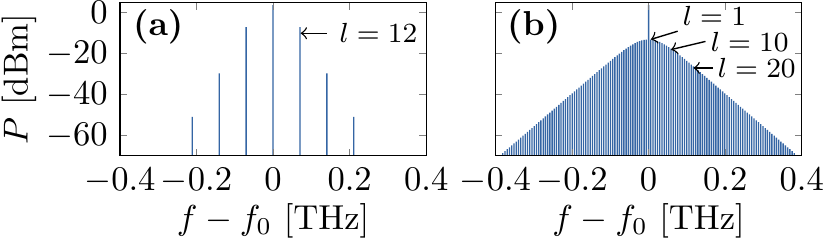}
  \caption{Simulated spectra of: 
   $(a)$ A Turing pattern with $L=12$.
   $(b)$ A cavity soliton.}
  \label{fig:Spectra}
\end{figure}

In order to investigate theoretically the phase-locking of Turing patterns and
their robustness relatively to solitons, an intermediate set-point was chosen,
with $\beta=-0.03$, $\alpha=0$, $F^2=2.5$ and noisy initial conditions, leading
to the generation of the $12$-th order Turing pattern shown in
Fig.~\ref{fig:Spectra}(a).  Contrary to Turing patterns which fill entirely the
cavity, the bright cavity soliton is a unique structure spinning in the
cavity~\cite{Coillet2013}. The spectrum of the soliton is therefore composed of
spectral lines separated by only one FSR, as shown on Fig.~\ref{fig:Spectra}(b)
[$\alpha=3$, $\beta = -0.03$, $F^2=3$; noisy Gaussian pulse initial condition].
Cavity solitons are known to be sub-critical dissipative structures, and in the
LLE they appear for values of $\alpha$ greater than $41/30$. 

Turing patterns and cavity solitons are the two most promising Kerr comb
regimes for ultra-pure microwave generation. They are indeed structurally stable
solutions of the LLE and can be obtained with parameters values far from chaotic
regimes. In this application, the microwave signal corresponds to the beating
between the different modes of the comb detected by a fast photodiode. In order
to evaluate the phase properties of this beat-frequency, one needs to investigate
the relative phases of the different modal components of the comb. 
From an analytical point of view, this task can hardly be performed using the LLE.
Instead, the following modal decomposition 
\begin{equation}
  \psi (\theta, \tau) = \sum_l \Psi_l (\tau) e^{i l \theta}
  \label{eq:expansion}
\end{equation}
allows to have direct access to the phase $\varphi (\tau)$ of the slowly-varying
modal amplitudes $\Psi_l = |\Psi_l | e^{i \varphi}$, with $l = \ell - \ell_0$
being the azimuthal eigennumber of the photons with respect to the pumped mode
(the pumped mode is therefore $l=0$, while the sidemodes correspond to $l = \pm
1, \pm 2, \dots$). By plugging the expansion of Eq.~(\ref{eq:expansion}) inside
the LLE, the following ordinary differential equations ruling the dynamics of
each mode is obtained:
\begin{align}
  \frac{d{\Psi}_l}{d\tau} =& 
  \left[- ( 1+ i \alpha) + i \frac{\beta}{2} \, l^2 \right] \Psi_l + \delta(l) \, F  \nonumber \\
              & + i \sum_{m,n,p} \delta(m-n+p-l) \, \Psi_m \Psi_n^* \Psi_p  \, ,
\label{eq:modalequations}
\end{align}
where $\delta (x)$ is the usual Kronecker function, while $m$, $n$, $p$, and $l$
are eigennumbers labelling the interacting modes following the interaction
$\hbar \omega_m + \hbar \omega_p \leftrightarrow \hbar \omega_n + \hbar
\omega_l$.  It is important to note that strict equivalence between the modal
and spatiotemporal models had been demonstrated in ref.~\cite{PRA_Yanne-Curtis},
so that the theoretical analysis can be performed with either formalism
depending on their specific advantages.

\begin{figure}
  \centerline{
    \includegraphics{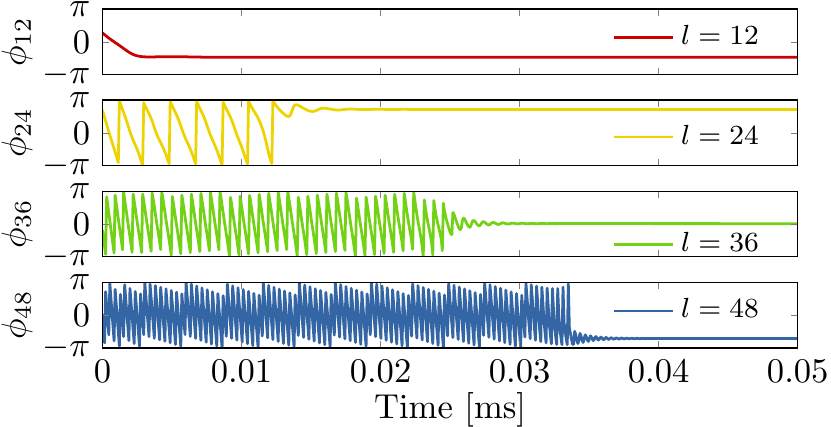}
  }
  \caption{Time evolution of the relative phases of the 4 first modes of the Turing
  pattern of Fig.~\ref{fig:CompExpSim}(a). After a delay increasing with the
  mode number, the relative phases reach a constant value, the Kerr comb
  becoming phase-locked.}
  \label{fig:PhaseTransTP12}
\end{figure}

The regime of Turing pattern is essentially originating from modulational
instability, which was analyzed in detail in
refs.~\cite{YanneNanPRA,YanneNanPRL,Coillet2013,wabnitz,lamont}. Turing patterns
actually arise from the following cascade of interactions.  When the pump is set
above the threshold $F^2_{\rm th} = 1+ (1- \alpha)^2$,  the modes around $\pm
l_{\rm th}$ with $l_{\rm th}= [(2(\alpha -2) /\beta)]^{1/2}$ experience positive
gain and are thereby excited.  After a transient dynamics characterized by modal
competition, two symmetric sidemodes $\pm L \simeq \pm l_{th}$ are stabilized
around the pump and the net flux of photons follows the degenerate  interaction
$2 \hbar \omega_0 \rightarrow \hbar \omega_{L} + \hbar \omega_{-L}$.  From then,
the sidemodes $\pm L$ provide in their turn a net flux of photons to the
sidemodes $\pm 2L$ using interactions $\hbar \omega_0 +  \hbar \omega_{\pm L}
\rightarrow  \hbar \omega_{ \mp L} + \hbar \omega_{\pm 2L}$.  This cascaded
process is further generalized later through the interactions $\hbar \omega_0 +
\hbar \omega_{\pm L} \rightarrow  \hbar \omega_{ \mp kL} + \hbar \omega_{\pm
(k+1) L}$ where the sidemodes $\pm kL$ provide a net flux of photons to the
sidemodes $\pm (k+1)L$.  This cascaded fountain effect asymptotically yields a
Kerr comb characterized by a multiple-FSR spacing of multiplicity $L$.  The
effect of this cascading mechanism can be observed in
Fig.~\ref{fig:PhaseTransTP12}, displaying the phase dynamics of the $12$-th
order Turing patterns of Fig.~\ref{fig:Spectra}(a).  It can be clearly seen that
after a transient dynamics, the various modal phases $\phi_l= \varphi_l -
\varphi_0$  with respect to the pump are reaching a time-independent value.  The
phase stabilization occurs first for the mode $l=12$, and then to $l=24$,
$l=36$, and so on. 

\begin{figure}
  \centerline{
    \includegraphics{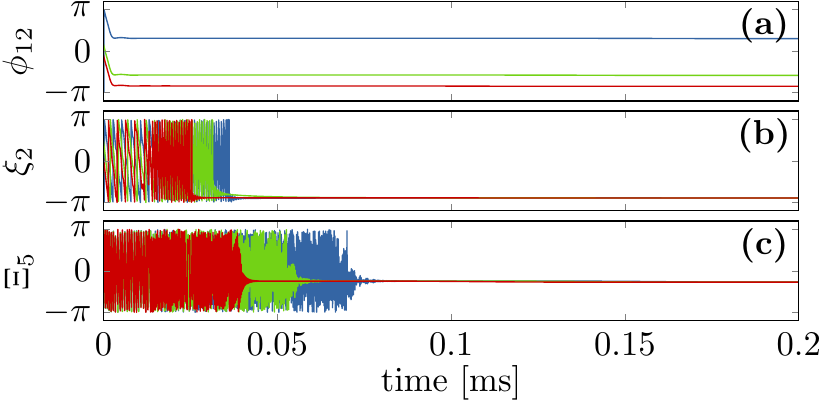}
  }
  \caption{Phase-locking behavior for different noisy initial conditions. (a) Phase
    difference $\phi_{12}= \varphi_{12} - \varphi_0$.  Evidence of triplet
    phase-locking ((b), $\xi_{2}$ is constant) and global phase-locking ((c),
    $\Xi_5$ is constant).} 
    \label{fig:PhaseTripletglobal}
\end{figure}

It is important to note that the phase locking value depends on the initial
condition, as illustrated on Fig.~\ref{fig:PhaseTripletglobal}(a), where the 
phase difference $\phi_{l}= \varphi_{l} - \varphi_0$ for $l=12$ has been plotted
for different initial conditions. This phenomenon is true for all the excited
modes of the Turing pattern.  However, it can be shown that an intrinsic 
phase-locking relationship takes place between the modes of the comb.
According to the scenario described earlier, a given mode $kL$ mainly interacts
with its two neighboring modes $(k \pm 1) L$.  In the steady state
($d\Psi_{kL}/d\tau \equiv 0$), this three-mode interaction can be rewritten as 
\begin{eqnarray}
\kappa_{kL} \Psi_{kL} & = &  \delta(k) \, F + |\Psi_{kL}|^2 \Psi_{kL} \nonumber \\
                      &&  2 \{ |\Psi_{(k-1)L}|^2 \Psi_{kL}   + |\Psi_{(k+1)L}|^2 \Psi_{kL} \nonumber  \\
                      && +  \Psi_{(k-1)L} \Psi_{kL}^* \Psi_{(k+1)L} \}
\label{eq:fixedpoint}
\end{eqnarray}
with $\kappa_{l} = - ( 1+ i \alpha) + i \beta l^2/2 $.  By extracting the phases
in this relashionship, it appears that consecutive triplets of modes are
necessarily phase-locked according to 
\begin{eqnarray}
\xi_k = \varphi_{(k-1)L} -2\varphi_{kL}+ \varphi_{(k+1)L}  =  {\rm Constant} \, . 
\label{eq:tripletphaselock}
\end{eqnarray}
Hence, the phases $\varphi_{kL}$ might individually have an asymptotic value
that depends on the initial conditions, but for the same Turing pattern, the
combination $\varphi_{(k-1)L} -2\varphi_{kL}+ \varphi_{(k+1)L}$ will be
\emph{independent} of initial conditions and will converge to the same value
$\xi_k$, as it can be seen in Fig.~\ref{fig:PhaseTripletglobal}(b).  The value
of  $\xi_k$ depends only on the parameters $\alpha$, $\beta$, $F$ and $k$.  The
sequence is therefore the following: initially, the triplet $\{-L,0,L \}$ is the
first to phase-lock. Then the two triplets $\{0, \pm L,\pm 2L \}$ phase lock,
and later on, all the cascade of triplets  $\{\pm (k-1)L, \pm kL, \pm (k+1)L \}$
as $k \rightarrow + \infty$ successively phase lock as well. 

\begin{figure*}[!th]
  \centerline{
    \includegraphics{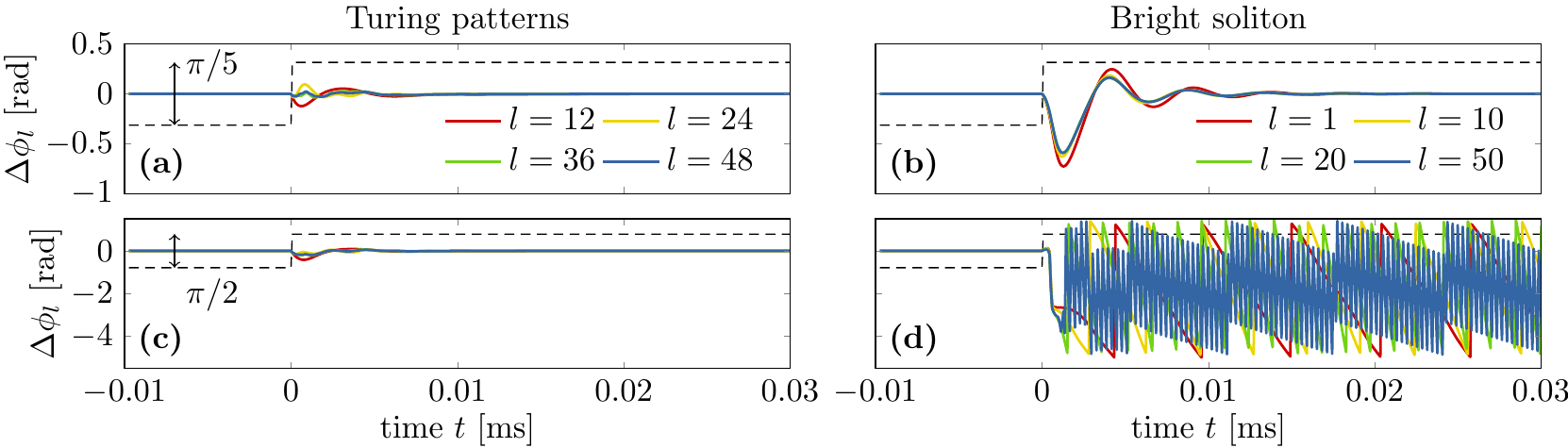}
  }
  \caption{Variation of the relative phases $\Delta \phi_l$ of different
    spectral modes of the Kerr combs while the pump signal undergoes an abrupt
    phase transition. (a) and (c): Case of the Turing pattern of
    Fig.~\ref{fig:Spectra}(a).  (b) and (d): Case of the bright soliton comb
    of Fig.~\ref{fig:Spectra}(b).  At $t=0$, the phase of the pump is shifted by
    $\pi/5$ in (a) and (b) and by $\pi/2$ for (c) and (d).  In the
    $\pi/5$ phase shift case, oscillations in the relative phases of the comb's
    lines are observed, but the phase-locking returns to its previous state and
    value.  The amplitude and duration of these oscillations are much larger in
    the case of the bright soliton compared to the Turing patterns case.  In the
    $\pi/2$ phase shift case, while the phase locking of the Turing pattern is
    maintained, the soliton is destroyed, and thus the phase locking.}
  \label{fig:PhaseBumps}
\end{figure*}
This sequence of triplet phase-locking ultimately leads to a global phase
coupling through a cascaded process, and the global phase-locking relationship
can be obtained by considering that since any $\xi_k$ is constant, so is any
linear combination of them. In particular, if we consider $2N+1$ oscillating
modes in a Turing pattern comb (the pump and $N$ excited modes on each side), we
can show that, 
\begin{align}
  \Xi_N &= \frac{1}{2} \sum_{k=-(N-1)}^{k=N-1} [1 - (-1)^{N+k} ] \, \xi_k \nonumber \\
    &=  \varphi_{_{-NL}} + 2 \sum_{k=-(N-1)}^{k=N-1}(-1)^{N+k} \varphi_{_{kL}}
    +\varphi_{_{NL}} \nonumber \\
    &= {\rm Constant} \, . 
\label{eq:globalphaselock}
\end{align}
If we consider for example the $7$ oscillating modes that are visible in
Fig.~\ref{fig:Spectra}(a), this global phase locking relationship simply reads
$\Xi_3 = \varphi_{-36} -2\varphi_{-24}+2\varphi_{-12} -2\varphi_{0}  +
2\varphi_{12} -2\varphi_{24} +\varphi_{36} = {\rm Constant} $. This relationship
can include more modes if needed [for example the modes $l=\pm 48$, which are
indeed oscillating as it can be seen in Fig.~\ref{fig:PhaseTransTP12}(d)].
Hence, when the permanent state is reached, the linear combination $\Xi_N$ which
involves all the modal phases is a constant that is  independent of the initial
conditions, as evidenced by Fig.~\ref{fig:PhaseTripletglobal}(c). These
intrinsic phase relations $\xi_k$ and $\Xi_N$ are therefore responsible for the
shape of the Kerr comb in the Turing pattern regime.

In order to test the robustness of the phase-locking in both the Turing pattern
and the soliton regimes, we introduce a perturbation after the stationary state
is reached: the phase of the pump signal is abruptly shifted by constant offset,
and we monitor the difference $\Delta \phi_l(t) = \phi_l(t) -
\phi_l^{\text{st}}$, with $ \phi_l = \varphi_l -\varphi_0$.  When the offset in
the pump phase is set to $\pi/5$, the relative phases of the different modes
oscillate in response to this perturbation, but the phase-locking is preserved.
In the case of the Turing patterns, the phase oscillations are of small
amplitude with respect to the perturbation, and decrease when the mode number
increases, as if the perturbation was attenuated during its propagation to
higher order modes. In contrast, the amplitudes of the phase oscillations of the
soliton are larger than the initial stimulus, and do not substantially decrease
for higher order modes. This result indicates that Turing patterns are more
robust to external phase perturbations, and could lead to better phase-noise
performances.  To strengthen this claim, the phase-shift of the pump signal was
increased to $\pi/2$, and the resulting phase evolution is plotted in
Fig.~\ref{fig:PhaseBumps}(c) and~(d).  The Turing pattern behavior remains very
stable and identically phase-locked, while the soliton is destroyed (as well as
the phase-locking). 

In this work we have given an insight into the dynamics of phase-locking in a
Kerr frequency comb.  The fact that the photonic interactions are strongly
dominated by exchange of microwave  photons of energy $L\times \hbar \Omega_{\rm
FSR}$  is a direct consequence of the modulational instability, which generates
the highest gain for photons whose offset frequency relatively to the pump is
precisely $L \times \Omega_{\rm FSR}$.  We have analytically demonstrated that
phase-locking in Turing patterns is achieved through a cascaded phase-locking of
adjacent triplets, which ultimately leads to a global phase-locking relationship
that we have determined.  We have also shown that Turing patterns are more
robust than solitons with regards to external perturbations, thereby
demonstrating their stronger potential for ultra-stable microwave
generation.

Y. C. K. acknowledges financial support from the European Research Council (ERC)
through the project NextPhase.


\end{document}